\documentclass[prl,twocolumn,aps,showpacs]{revtex4}
\usepackage{graphicx}
\usepackage{amsmath, amssymb}

\newcommand{\figref}[1]{Fig. \ref{#1}}

\newcommand{\figwidth}{8.5cm}

\begin{document}

\title{Tapered-amplified AR-coated laser diodes for Potassium and Rubidium atomic-physics experiments}

\author{R. A. Nyman}
\email{robert.nyman@iota.u-psud.fr}
\homepage{http://atomoptic.iota.u-psud.fr}
\affiliation{Laboratoire Charles Fabry de l'Institut d'Optique, UMR8501 du CNRS, B\^{a}t. 503, Centre Scientifique d'Orsay, 91403 Orsay Cedex, FRANCE}

\author{G. Varoquaux}
\affiliation{Laboratoire Charles Fabry de l'Institut d'Optique, UMR8501 du CNRS, B\^{a}t. 503, Centre Scientifique d'Orsay, 91403 Orsay Cedex, FRANCE}

\author{B. Villier}
\affiliation{Laboratoire Charles Fabry de l'Institut d'Optique, UMR8501 du CNRS, B\^{a}t. 503, Centre Scientifique d'Orsay, 91403 Orsay Cedex, FRANCE}

\author{D. Sacchet}
\affiliation{Laboratoire Charles Fabry de l'Institut d'Optique, UMR8501 du CNRS, B\^{a}t. 503, Centre Scientifique d'Orsay, 91403 Orsay Cedex, FRANCE}

\author{F. Moron}
\affiliation{Laboratoire Charles Fabry de l'Institut d'Optique, UMR8501 du CNRS, B\^{a}t. 503, Centre Scientifique d'Orsay, 91403 Orsay Cedex, FRANCE}

\author{Y. Le Coq\footnote{Present Address: Optical Frequency Measurement Group, Time and Frequency Division (MS 847.10), NIST, 325 Broadway, Boulder CO 80302, USA}}
\affiliation{Laboratoire Charles Fabry de l'Institut d'Optique, UMR8501 du CNRS, B\^{a}t. 503, Centre Scientifique d'Orsay, 91403 Orsay Cedex, FRANCE}

\author{A. Aspect}
\affiliation{Laboratoire Charles Fabry de l'Institut d'Optique, UMR8501 du CNRS, B\^{a}t. 503, Centre Scientifique d'Orsay, 91403 Orsay Cedex, FRANCE}

\author{P. Bouyer}
\affiliation{Laboratoire Charles Fabry de l'Institut d'Optique, UMR8501 du CNRS, B\^{a}t. 503, Centre Scientifique d'Orsay, 91403 Orsay Cedex, FRANCE}


\begin{abstract}
We present a system of room-temperature extended-cavity grating-diode lasers (ECDL) for production of light in the range 760--790nm. The extension of the tuning range towards the blue is permitted by the weak feedback in  the cavity: the diodes are anti-reflection coated, and the grating has just 10\% reflectance. The light is then amplified using semiconductor tapered amplifiers to give more than 400mW of power. The outputs are shown to be suitable for atomic physics experiments with potassium (767nm), rubidium (780nm) or both, of particular relevance to doubly-degenerate boson-fermion mixtures.
\end{abstract}


\date{\today}
\maketitle

\section{Introduction}

Atomic-physics experiments have strict demands on the quality of the laser systems used for trapping and cooling of atoms. One of the most popular atomic species is $^{87}$Rb which, for instance, can be easily cooled to quantum degeneracy \cite{BEC}. It has a cycling transition close to 780nm \cite{Wallace92}, used for cooling and trapping, e.g. optical molasses and magneto-optical traps. For this wavelength, semiconductor laser-diode systems have been
available for some time, due to their use in compact disc players and recorders. More recently, mixing different atomic species in trapping and cooling experiments \cite{Santos95,Shaffer95}, in particular $^{87}$Rb and $^{40}$K  \cite{Cataliotti98,Prevedelli99}, has encountered renewed interest with the studies of Bose-Fermi degenerate
mixtures \cite{DeMarco99}. The cycling transition which can be used for cooling and trapping potassium is at 767nm  \cite{Williamson95}, for which semiconductors have only recently become available. So far, most groups working with atomic potassium produce light at this wavelength using either Titanium-Sapphire lasers \cite{Cataliotti98,Prevedelli99} or by cooling diode lasers designed for 780nm below zero degrees  Celsius\cite{DeMarcoThesis,Fletcher04}. In both cases, the result is a complex, and sometimes expensive, set-up.

The use of diode lasers for atom-cooling experiments has been made possible by the development of extended-cavity diode lasers (ECDL) \cite{Wieman91}. These take advantage of the available laser diodes and use frequency-selective feedback and high photon numbers to achieve tunability and narrow linewidth, typically via a diffraction grating in either the Littrow \cite{Arnold98,Cassetari98} or Littman configuration \cite{Littmann78}. Using the temperature dependance of the semiconductor gain medium has allowed use of mass-produced diode lasers, whose room temperature wavelength is controlled to within a few nanometres, for a rich variety of atomic species. In particular, diode lasers centred around 785nm are commonly used for rubidium trapping (780nm) when used at room temperature and for the 767nm potassium transition, provided they are then cooled to sub-zero temperature. Consequently, diode-lasers so far developed which present the capability of tuning across the full spectra of the two species without significant changes in temperature, alignment, or even the laser diode are only of low power, suitable for spectroscopy\cite{Banerjee04}. Since ECDLs do not usually deliver more than a few mW of useful light, amplification is required for atomic trapping and cooling, by injection locking \cite{Wieman91}, through a broad-area laser \cite{Shvarchuck00} or via a tapered amplifier  \cite{Voigt01,Goldwin02,Aubin05}. An alternative is to put the high-power semiconductor element in cavity, resulting in either the tapered laser (a commercial example is used in ref. \cite{Catani05}), or an external-cavity broad-area laser diode\cite{Cassetari98}, which have the drawback of requiring complex mechanical engineering, cooling and beam-shaping.

\begin{figure}
    \centering
  \includegraphics[width=\figwidth]{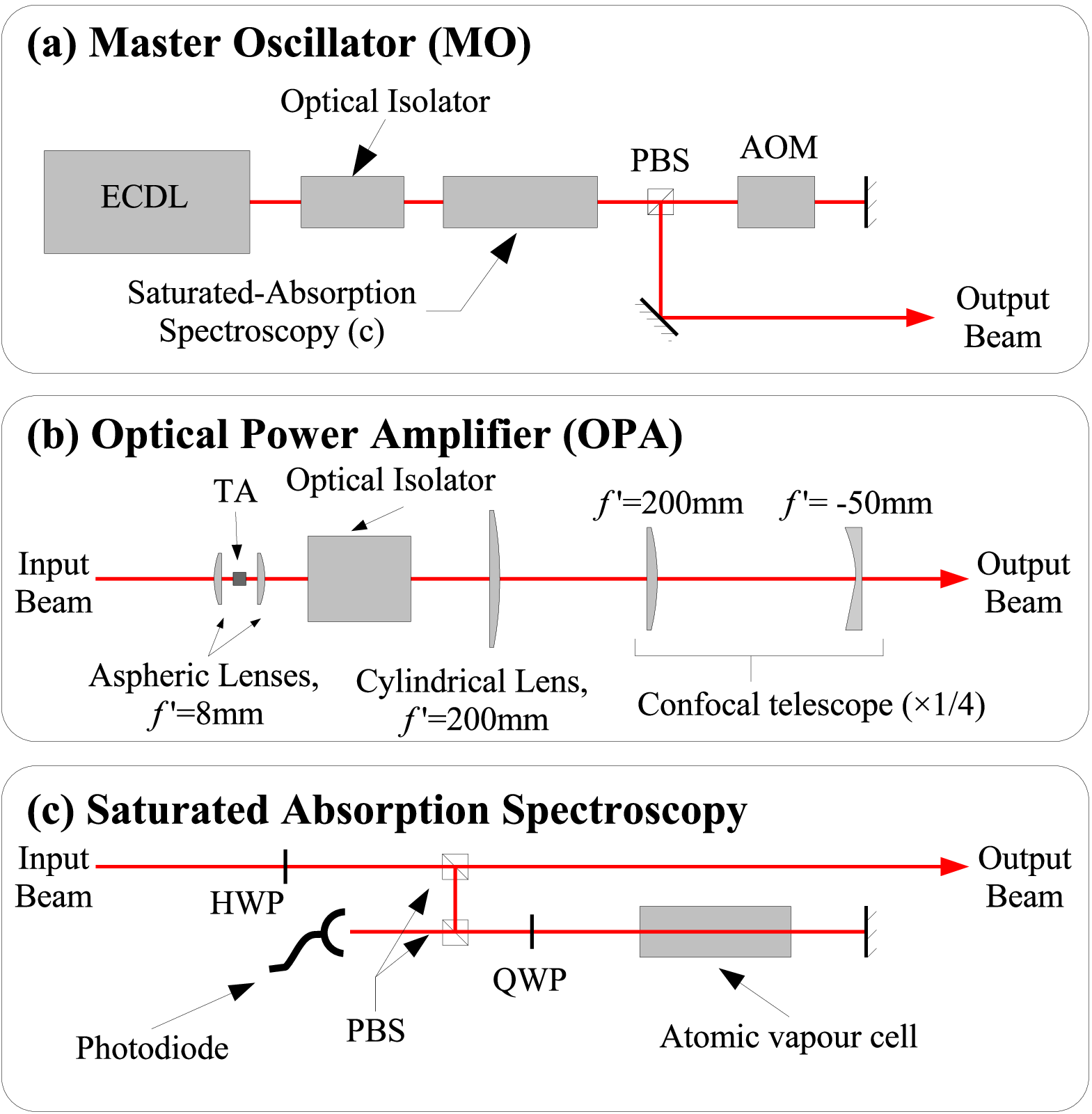}\\
  \caption{Schematic of apparatus for the production of light of the power
  and wavelength required. Abbreviations used: ECDL (extended-cavity diode laser), PBS (polarising beam-splitter cube), AOM (acousto-optical modulator in double-pass configuration), TA (tapered-amplifier chip), HWP (half-wave plate), QWP (quarter-wave plate).}
  \label{fig:optical schema}
\end{figure}

In this paper, we present a system of Master Oscillators (MOs) and Optical Power Amplifiers (OPAs, see \figref{fig:optical schema}) which can be tuned anywhere in the range 760--790nm close to room temperature, thus being useful for simultaneous cooling and trapping of Rb and K. The master oscillators in our system are Littrow-type ECDLs using antireflection-coated laser diodes. The wavelength of the light is locked to an atomic transition using saturated-absorption spectroscopy, then shifted using an acousto-optical modulator (AOM). The AOM permits
rapid control of the amplitude of the MO beam. The power output from the cavity can be up to 50mW, more than 50\% of the nominal power of the free running laser (when not AR coated). One or more MO beams form the input to an Optical
Power Amplifier (\figref{fig:optical schema}), which is a semiconductor tapered amplifier with a gain range similar to that of the laser diodes. After appropriate beam shaping, the output of the amplifier can be injected into one or more single-mode optical fibres with typical efficiency up to 60\%.

\section{Master Oscillators: ECDL with Anti-reflection-coated Diode Lasers}

\subsection{Construction}
\begin{figure}[htb]
    \centering
  \includegraphics[width=\figwidth]{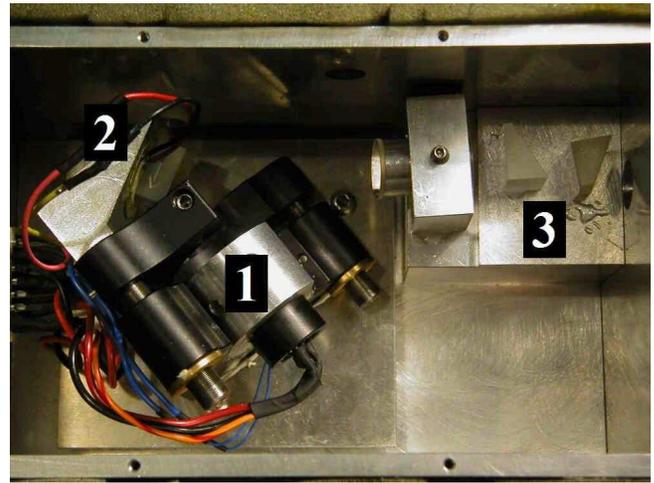}\\
  \caption{Picture of the ECDL (Littrow configuration). (1) Diode laser mount (\textit{Thorlabs} LT240P-B) which also acts as a mount for the collimation lens. (2) Holographic grating. (3) Anamorphic Prisms.}
  \label{fig:diode laser}
\end{figure}

The ECDLs follow the design of Arnold et al \cite{Arnold98} modified by C. Aussibal \cite{AussibalThesis} (see
\figref{fig:diode laser}). Briefly, the diode is mounted in a collimation tube (\textit{Thorlabs} LT240P-B) with an
aspheric collimation lens. The tube is then mounted into a modified, high-precision mirror-mount, which is held on a thermoelectric cooling device (TE cooler). A holographic diffraction grating is glued onto a piezoelectric element which is itself fixed on the moving part of the mirror mount, putting the ECDL in the Littrow configuration: the first diffraction order is retro-reflected and the zero order is output. The orientation of the grating with respect to the diode can be changed by adjustment of the modified mirror mount. The length of cavity can be finely adjusted by altering the voltage across the piezoelectric element.

Most ECDL designs employ standard diode lasers \cite{Wieman91} which thus create a coupled-cavity system.
This coupling can result in collapse of coherence \cite{Sacher92} which increases the linewidth of the laser,
and can make the mode-hop-free continuous scanning range much less than the free-spectral range of the extended cavity. In our ECDLs, we use \textit{Eagleyard}, Ridge-Waveguide Laser (SOT03 package), nominally specified for laser emission at 790nm (reference EYP-RWE-0790-0400-0750-SOT03-0000). The diodes have been anti-reflection coated on the output facet, to about 0.1\% reflectance, which broadens the gain spectrum of the semiconductor medium. To understand this, one notes that in diode laser heterojunctions, the region where free electrons and holes exist simultaneously, the active region, is confined to a thin middle layer. The light is also confined to this region, where the amplification takes place, leading to a high photon density in the junction. Above the laser threshold the carrier density is given by the balance between the injection of charge carriers due to the current through the junction, and the loss of carriers through stimulated radiative transitions. Reducing the optical intensity in the junction allows for an increased carrier population, and therefore an increased gap between the quasi-Fermi levels \cite{Chow94}.
This increased band gap leads to an enhanced gain in the blue end of spectrum. When AR-coated diodes are used in ECDLs, the intra-cavity intensity  for a given electrical current is indeed reduced in comparison with a non-antireflection-coated diode laser. Thus the total tuning range of diode lasers is extended to the blue\cite{Hildebrandt03}.

The choice of the grating is crucial for optimal operation. Compared to standard ECDLs, since there are no coupled-cavity dynamics, the cavity mode is stable for a lower grating-reflectivity in our system. Low finesse is required to avoid high in-cavity intensity, which could damage the semiconductor.In particular, the grating has to be
chosen so as best to match the reflectance of the surface of a non-AR-coated diode laser, namely about 7\%. We chose for a (cheap) grating a 1200 lines/mm holographic grating, optimised for UV light (\textit{Edmund Optics} T43-772), which diffracts roughly  10\% of light into the first order, and sends the remaining light into the zero-order output beam. The orientation of the grating with respect to the diode laser controls the wavelength and optimises the output intensity, by diffracting the first-order light into the internal laser cavity, and reflecting the remaining light into the output beam (the zeroth order). Since the diode sits in a cavity of very  low finesse (around 1), the  intensity of light in the semiconductor element is scarcely greater than the intensity of the output beam.

The extended cavity has a free spectral range of roughly 5GHz (0.01nm). Thus continuous scans can be performed of up to 5GHz around a wavelength in the range 760--790nm, by changing the length of the cavity using the voltage across the piezoelectric element. The central wavelength is chosen by turning the screws on the mirror mount, adjusting the angle of the grating with respect to the diode.

\subsection{Performance}

The AR-coated diode laser is specified to operate with a typical free-running centre wavelength of 780nm at 25$^\circ$C ($d\lambda/dT \sim +0.3$nm/$^\circ$C), with a spontaneous emission spectrum between 750 and 790nm. For our 1200 line/mm grating, the angle between the grating and the optical axis of the cavity is 27.9$^\circ$ for 780nm and 27.4$^\circ$ for 767 nm, a range achievable by adjusting the mirror mount orientation. Although mode matching via fine-tuning the temperature is required, we found that the desired wavelengths can consistently be found between 18 and 25$^\circ$C.
Typical threshold current for the ECDL is 35mA. Operating at 90mA, the output power is about 40mW.

\begin{figure}[htb]
    \centering
  \includegraphics[angle=-90, width=\figwidth]{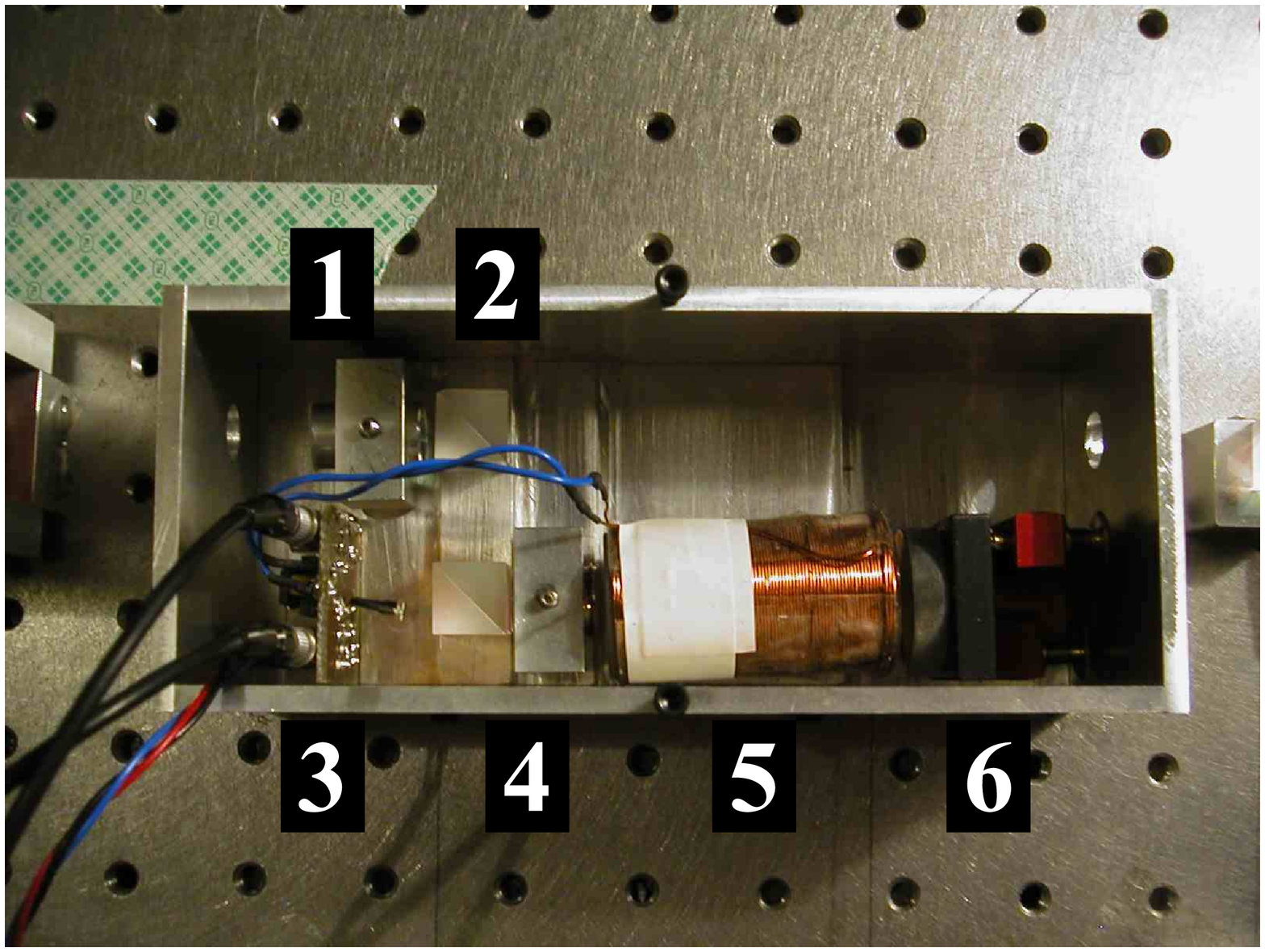}\\
  \caption{Saturated absorption spectroscopy module. (1) Half-wave plate.
(2) Polarising beam-splitter cube. (3) Photodiode. (4) Quarter-wave
plate. (5) Atomic vapour cell, surrounded by electromagnet. (6) Mirror.
The box is about 17cm long, by 7cm wide by about 6cm deep, and can be
closed to reduce background light.} \label{fig:sat abs photo}
\end{figure}

\begin{figure}[htb]
    \centering
  \includegraphics[width=\figwidth]{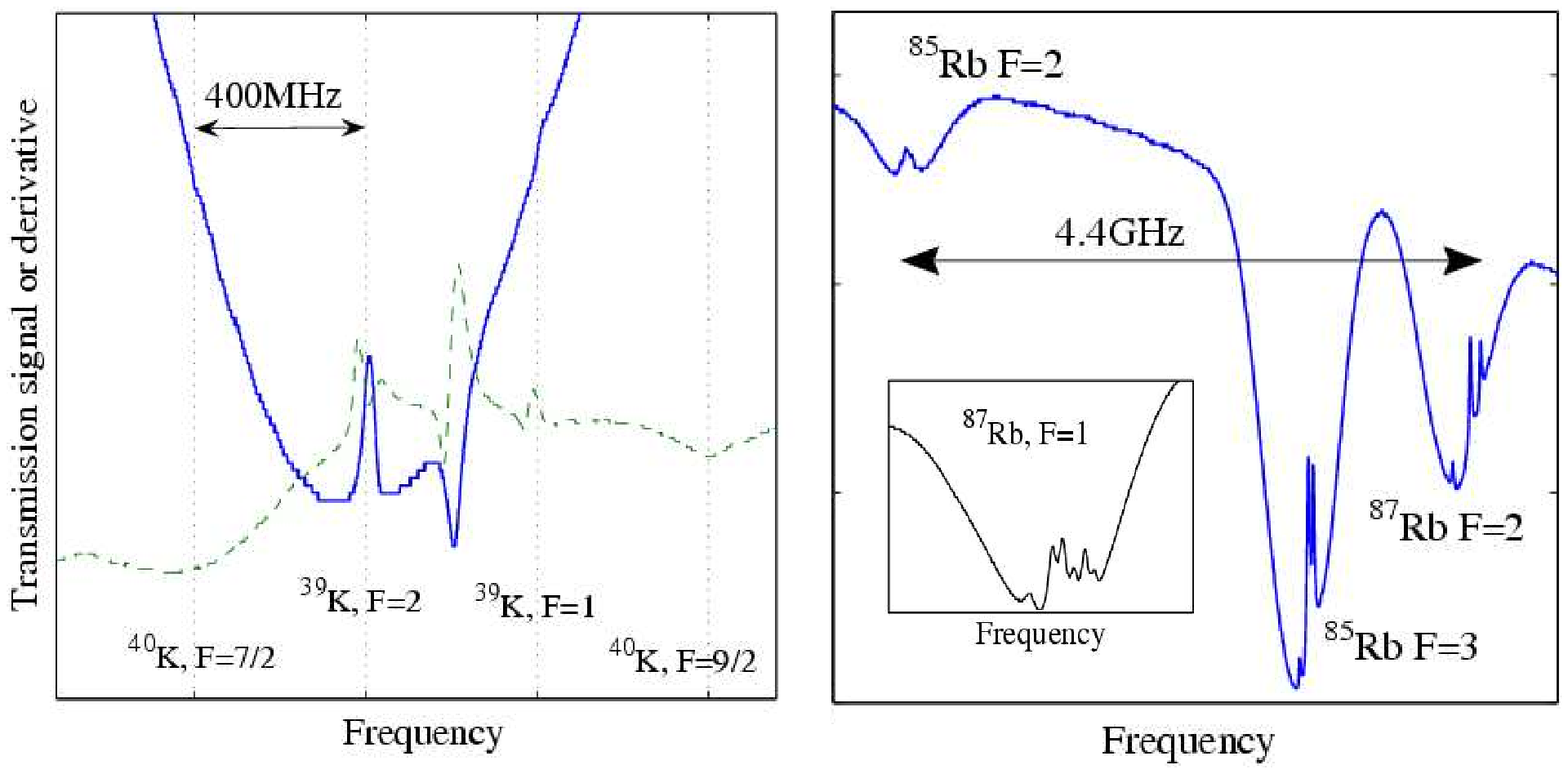}\\
  \caption{Saturated absorption spectra. On the left: the spectrum of
natural abundance potassium (93\% $^{39}$K), heated to about 60$^\circ$C. The dashed curve is
derivative of the saturated-absorption signal (solid curve). On the
right: Natural abundance rubidium saturated-absorption signal (at room temperature). The free
spectral range of the extended cavity is 5GHz. Inset: narrow scan of
$^{87}$Rb, $|F=1\rangle \rightarrow |F'=0,1,2\rangle$ transitions. Each
peak has a natural line-width of 6MHz.} \label{fig:sat abs}
\end{figure}

For use in laser cooling, the MOs are frequency-locked to the corresponding atomic transitions using sub-Doppler saturated-absorption spectroscopy \cite{satabs}. The spectroscopy module is designed to be compact and robust (\figref{fig:sat abs photo}) and allows monitoring of the spectra of natural-abundance potassium and rubidium as shown in \figref{fig:sat abs} (left and right respectively). The hyperfine lines have widths of order 6MHz for both species, but the excited state energy levels of potassium are too close to be resolved \cite{Santos95}. As expected, scans of more than about 5GHz are interrupted by mode hops of the extended cavity. For Zeeman-effect-modulation locking a magnetic field directed along the optical path and oscillating at 60kHz is applied, shifting the $m_F$ sub-levels. The error signal (derivative of spectroscopic signal) is generated using phase-sensitive detection, and then fed back via a proportional-integral loop to the master oscillator. The proportional signal is added to the laser diode current, and the integral signal to the voltage across piezoelectric element. The feedback loop has a bandwidth of roughly 1.5kHz.

\begin{figure}
    \centering
  \includegraphics[width=\figwidth,angle=0]{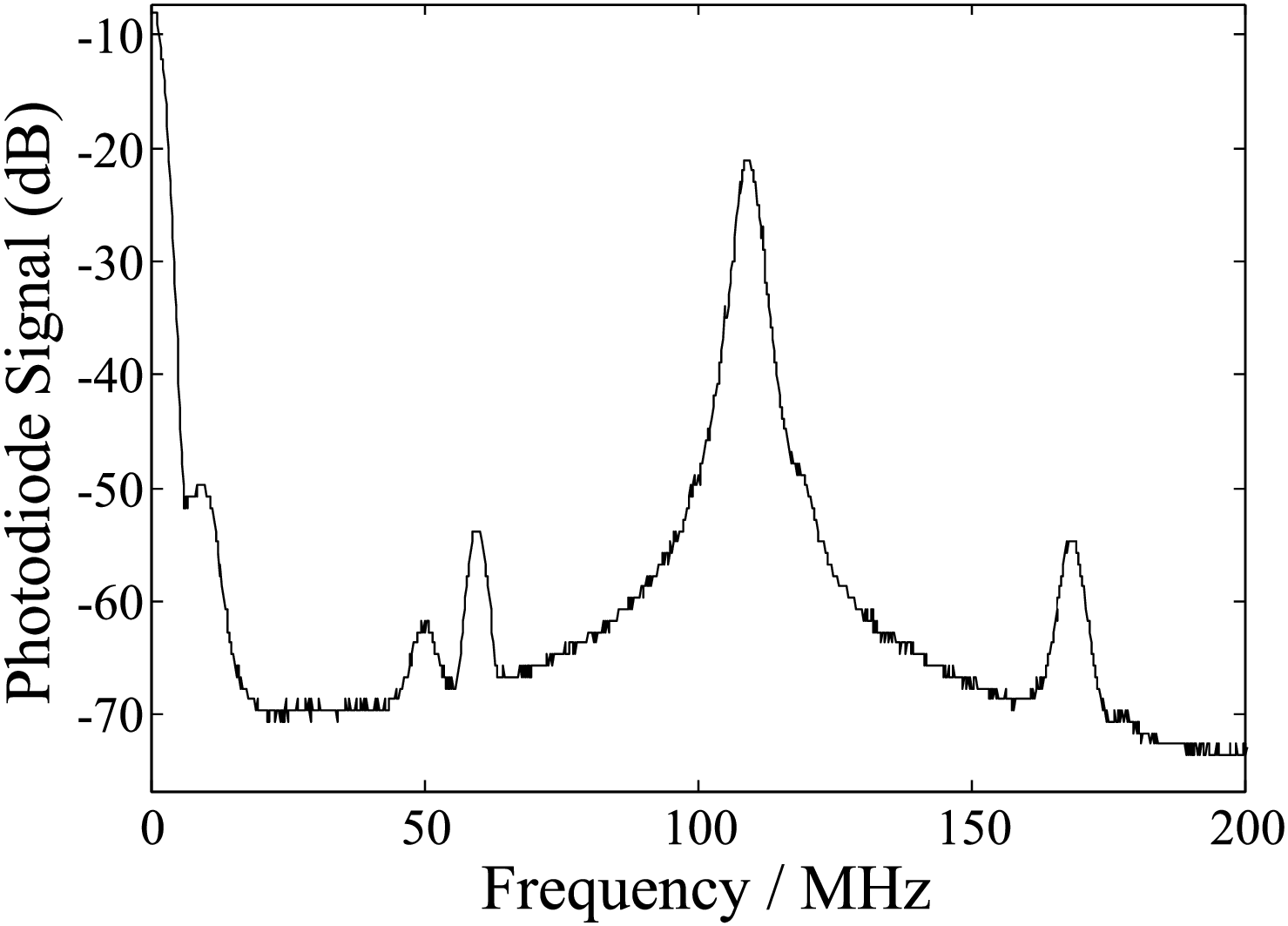}\\
  \caption{Beam-beating experiments, averaging of 64 measures during one
minute. The two beams are passed twice through acousto-optic modulators
at 59.5MHz and 114MHz (taking the first diffraction order each time),
giving a beat note at 109MHz, plus some weaker beat notes due to unwanted
diffraction orders in the double-pass AOM, or due to harmonics in the RF-driving electronics.}
  \label{fig:laser beats}
\end{figure}

We have measured the linewidth of the MOs using a heterodyne measurement between two systems at 780nm. The outputs of the two MOs were locked to the same atomic $^{87}$Rb transition, then shifted by a few MHz with AOMs, and superimposed on a fast photodiode. The beat-note of two non-AR-coated ECDLs (with gratings which reflect about 40\% into the first diffraction order) has a full width at half-maximum of 600kHz, i.e. each ECDL has a linewidth around 300kHz\cite{LeCoqThesis}, p39), if we assume that the lineshape is Lorentzian. The beat note of one AR-coated and one non-AR-coated ECDL, was recorded using a multi-channel spectrum analyser, as shown in \figref{fig:laser beats}. The full width at -30dB was $(22.3 \pm 0.2)$MHz, corresponding to a Lorentzian full-width at half maximum of $(706 \pm 6)$kHz. We therefore deduce that the frequency width of the ECDL with anti-reflection coated diode is around 400kHz. We conclude that  AR coating the diode and changing the grating have not significantly increased the linewidth. It is worth noting that in both cases, the linewidth of the laser is not limited by spontaneous emission (modified Schawlow-Townes limit \cite{Schawlow58,Henry82}), but by additional technical noise.

In terms of long term use, we find that the ECDLs can be kept frequency-locked for many hours, and locking is limited by external acoustic noise. Locking is consistently more stable on the potassium line since the
spectrum contains broader features, so the capture range is larger for K than for Rb. No thermal drifts of the ECDL cavity optic path length (including variation of refractive index in the semiconductor medium) were observed since the whole cavity is thermally stabilised, even though small drifts that occur when the temperature of the laboratory changes are not completely compensated for by the thermal feedback mechanism. Running continuously at between an output power between 30 and 50mW, for a total of about 42 diode-months (30,000 hours) we have seen no spontaneous failures of the laser diodes.

\section{Optical Power Amplifiers}

After proper frequency synthesis, using AOMs, the MOs are amplified. The optical power amplifiers
(OPAs) are tapered amplifiers \cite{Ferrari99} which offer significant
gain in the same wavelength range as the master oscillators, 760--790nm
(\textit{Eagleyard}, EYP-TPA-0780-00500-3006-CMT03, C-Mount 2.75mm package). The OPAs are mounted in a compact
self-contained mechanical housing (almost no adjustable parts) that allows very good temperature control of the chip and very good mechanical stability.

\subsection{Construction}

\begin{figure}
    \centering
    \includegraphics[width=\figwidth]{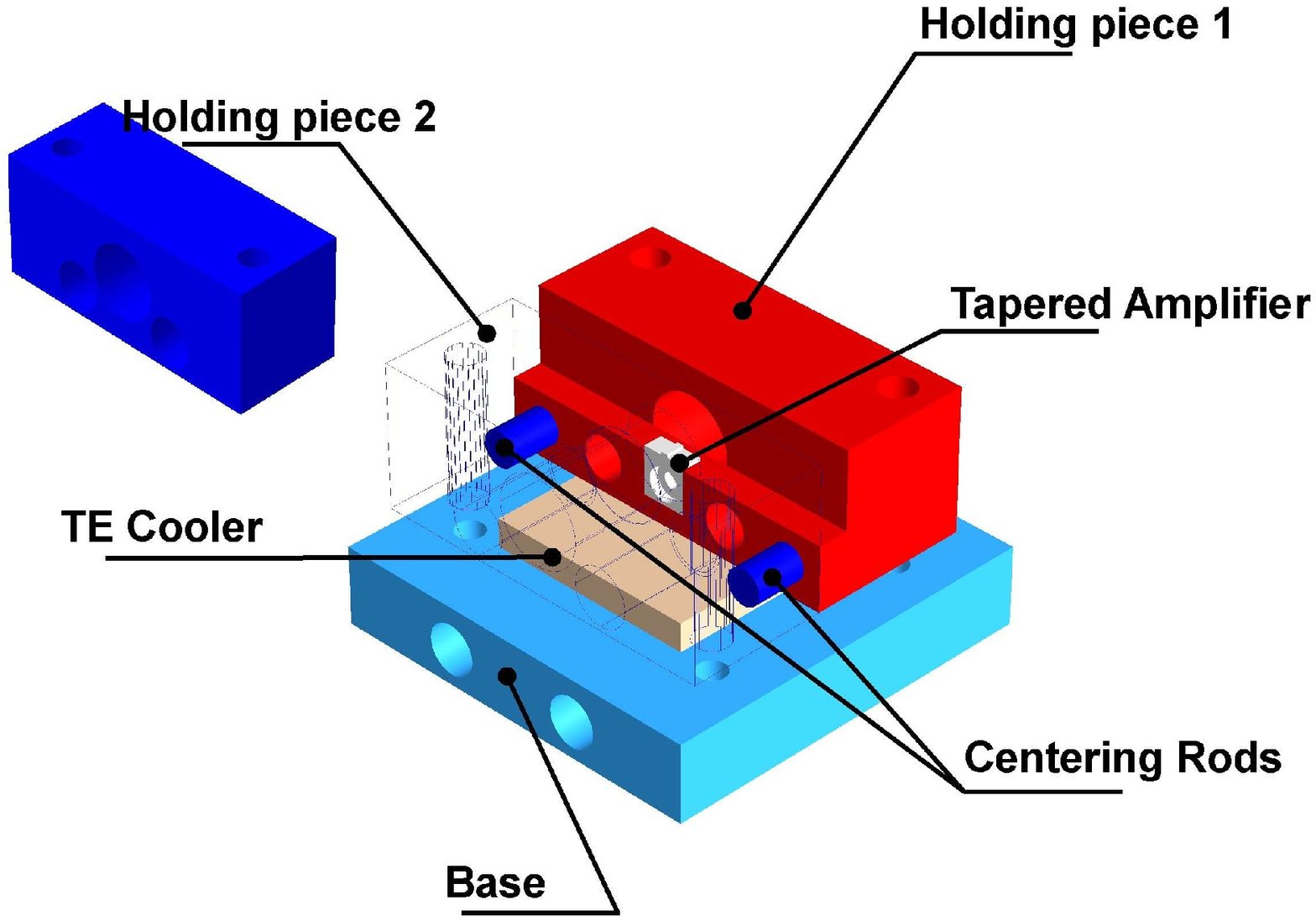}
    \includegraphics[width=\figwidth]{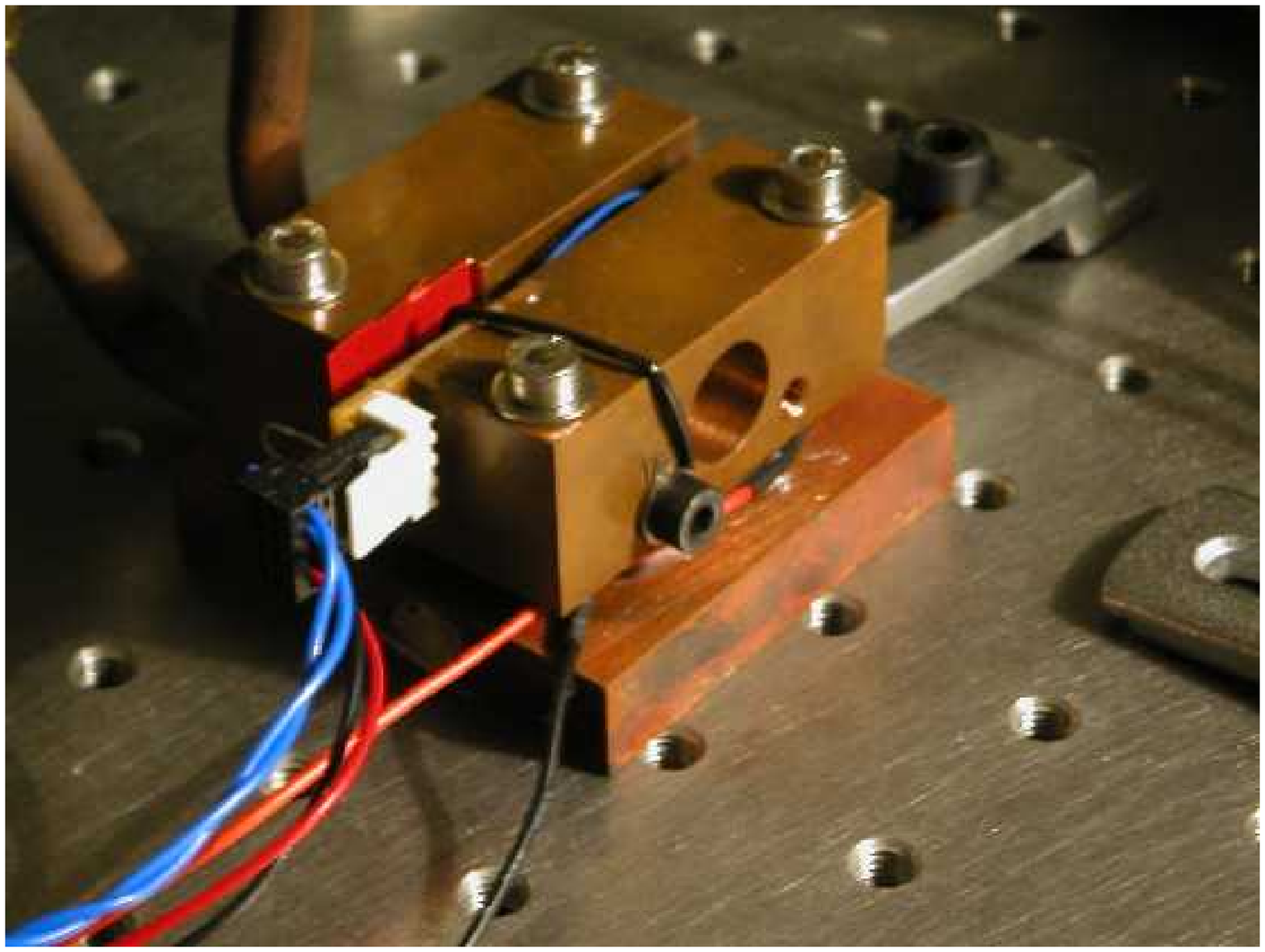}\\
    \caption{Mechanical and Thermal housing for the tapered amplifier (First-generation design).
Top:  Exploded diagram. The blocks are made of copper (the centring rods are of steel). Bottom: photograph of the assembled system.}
  \label{fig:Tapered Amplifier Mechanics}
\end{figure}

\begin{figure}
    \centering
  \includegraphics[width=\figwidth]{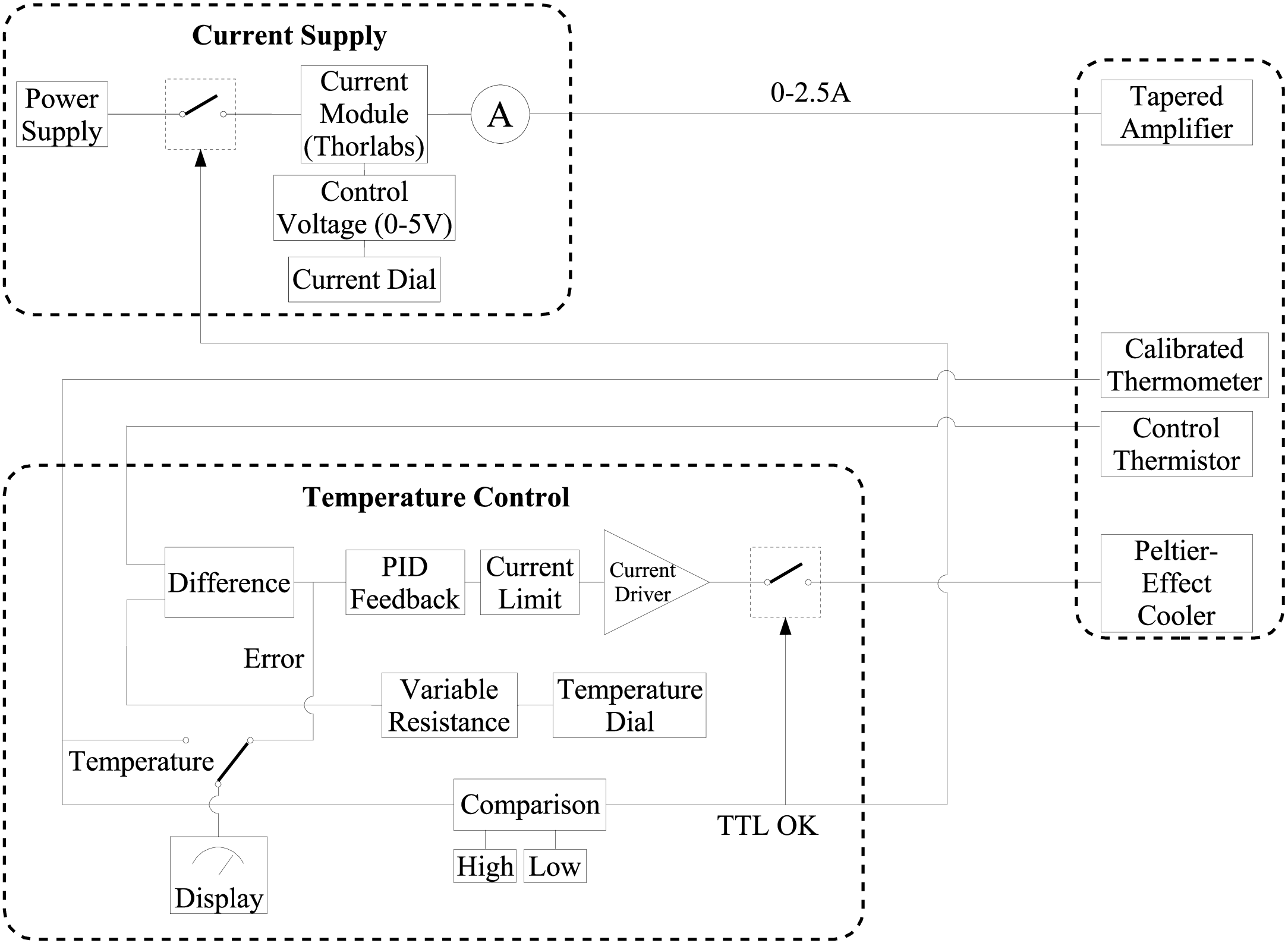}\\
  \caption{A Schematic diagram of the Electronics for the tapered
amplifier. If the temperature leaves a pre-defined range, the TTL-OK
signal is cut, and both temperature control and current supply are
switched off. The desired temperature and current are can be changed
during operation, using dials outside the electronics boxes. Other
adjustable parameters (e.g. PID gain, acceptable temperature) are
optimised before operation. See also \figref{fig:elec mopa temperature} and \figref{fig:elec mopa current}.}
  \label{fig:Tapered Amplifier Electronics}
\end{figure}

The mechanical supports hold the lenses to focus incoming light, to collimate the output light, and also fix the position of the tapered amplifier chip. Our compact design places the chip in the middle of a copper block, which helps control the chip temperature(see \figref{fig:Tapered Amplifier Mechanics}). Technical drawings of a second-generation design are available(\cite{technical}). Up to 10W of heat from the amplifier are dissipated to the water-cooled copper base plate via a TE-cooler. An un-calibrated thermistor is used as a temperature monitor by the temperature control unit (see \figref{fig:Tapered Amplifier Electronics}). Since the magnitude and wavelength variation of the gain vary significantly with temperature, the temperature of the amplifier is servo-locked. In addition, a calibrated thermometer is used for fault detection. The copper block also holds the contacts for the  amplifier current supply. Onto the base plate (of the second-generation design) can be bolted a lid so the chip can be held under vacuum at sub-zero temperatures, to shift the wavelength of maximal gain, if needed. A water-free environment is required to prevent condensation on the chip when it is cooled.

\begin{figure}[p]
    \centering
    \includegraphics[angle=90,width=\textwidth]{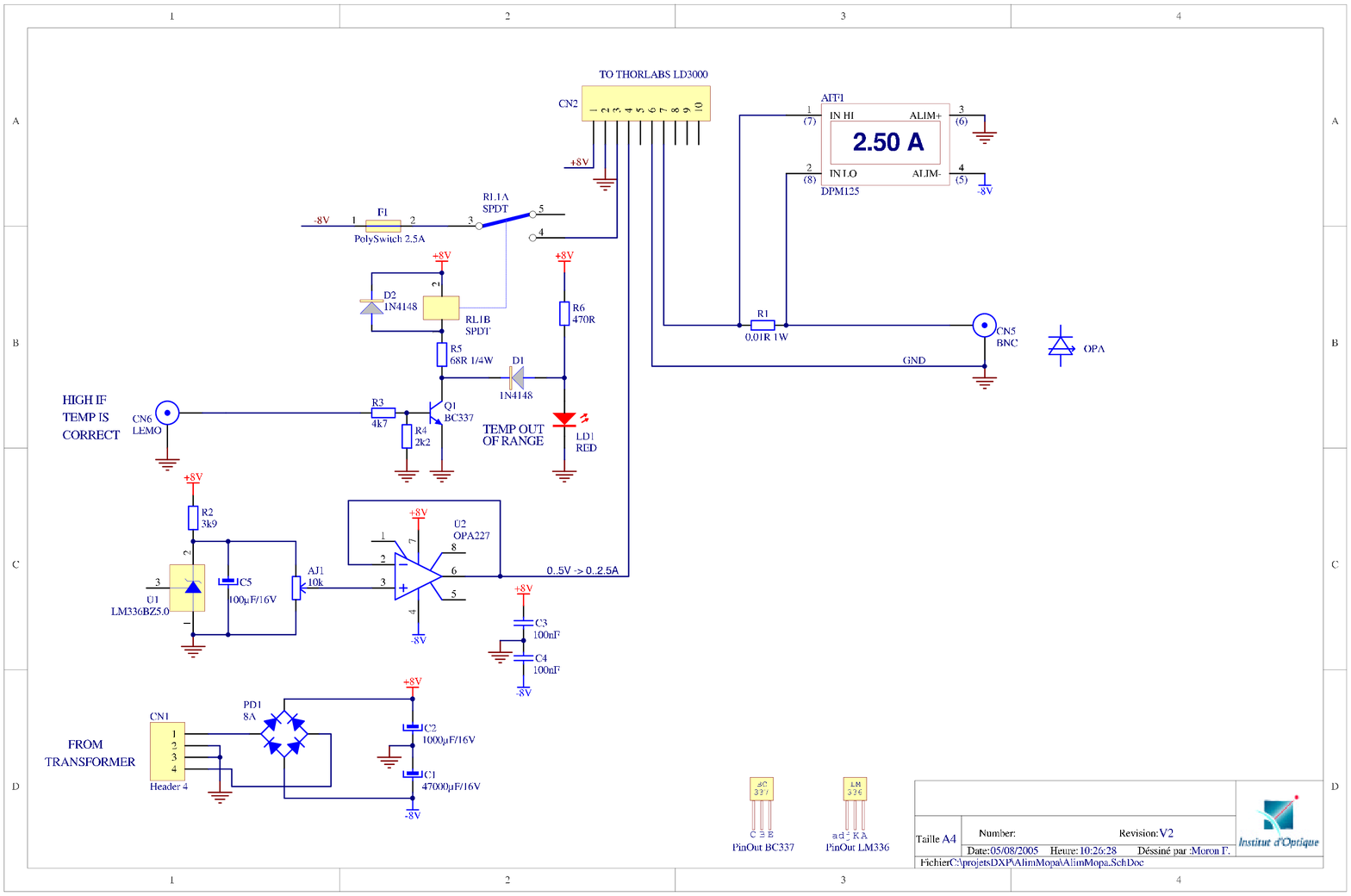}\\
    \caption{Electronic circuit diagram for the current supply unit for
tapered amplifier.} \label{fig:elec mopa current}
\end{figure}

The electronic systems for the OPAs consist of home-made boxes current supply and the temperature stabilisation, as shown schematically in \figref{fig:Tapered Amplifier Electronics}. Full electronic circuit diagrams are given in  and \figref{fig:elec mopa current}. The power supply (\figref{fig:elec mopa current}) is based around a low-noise, 2.5A current-supply module (\textit{Thorlabs} LD3000). The temperature-control unit (fig. \figref{fig:elec mopa  temperature}) is based around a proportional-integral-derivative feedback loop, the output of which drives the TE-cooler. If the temperature of the chip is outside a defined range, both units shut down, leaving the block to return slowly to ambient temperature; the current-supply unit has a soft-stop mechanism to protect the amplifier.

The output of the tapered amplifier is highly divergent and astigmatic.
One axis is collimated using the lenses built into the copper mounting
block; the other axis is corrected using a cylindrical lens. The beam is
then passed through a telescope to produce a collimated beam of around 2mm
diameter, then sent through an optical isolator. The light can be
injected in into one or more single-mode optical fibres with up to 60\%
efficiency.

\subsection{Performance}

\begin{figure}
    \centering
  \includegraphics[width=\figwidth]{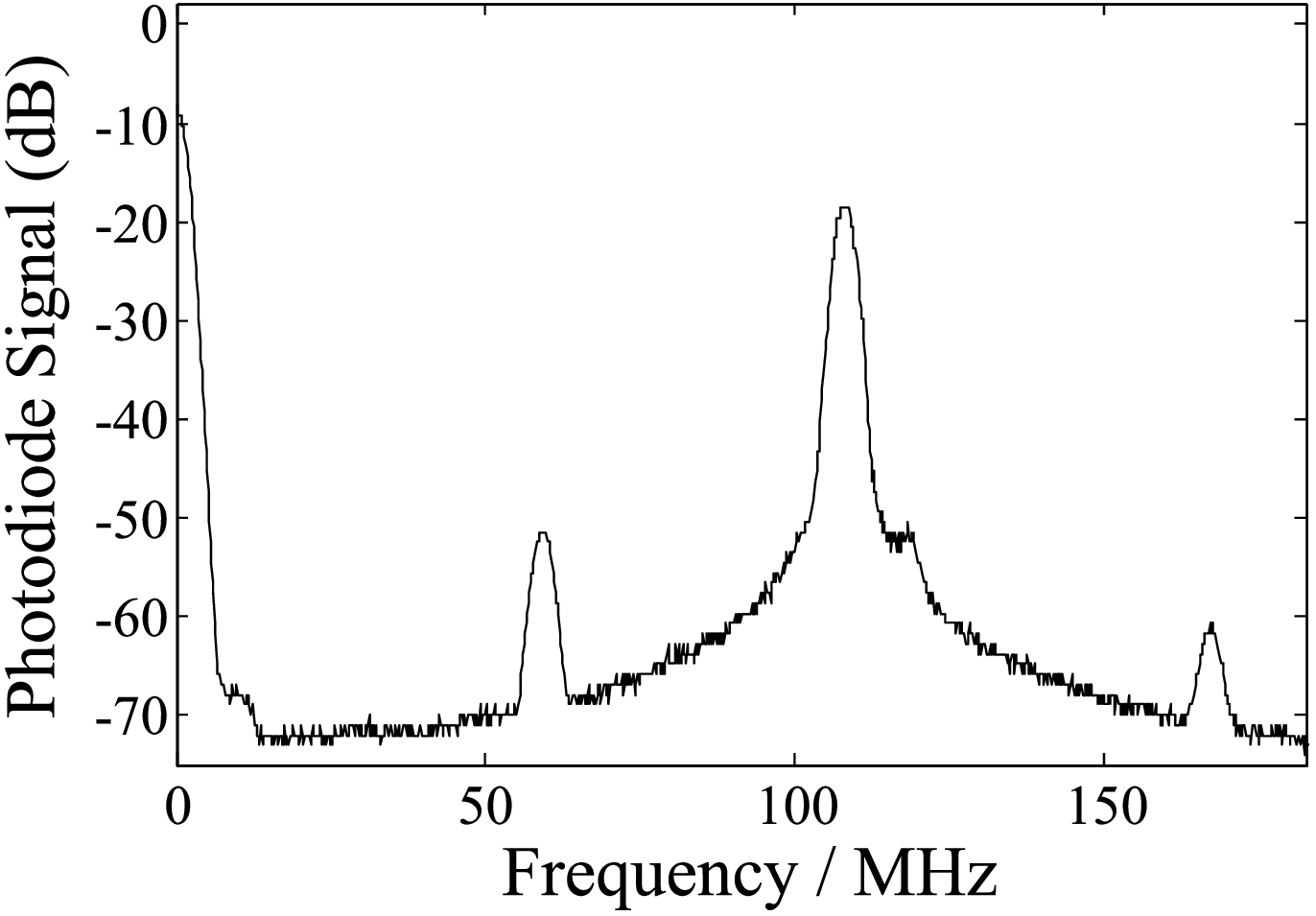}\\
  \caption{Modulation of output-beam of tapered amplifier. The same two
beams used for the demonstration of the laser line-width
\figref{fig:laser beats} were amplified, and the signal recorded. The
spectrum is the average of 64 measurements during one minute.}
\label{fig:mopa beats}
\end{figure}

\begin{figure}
  \includegraphics[width=\figwidth]{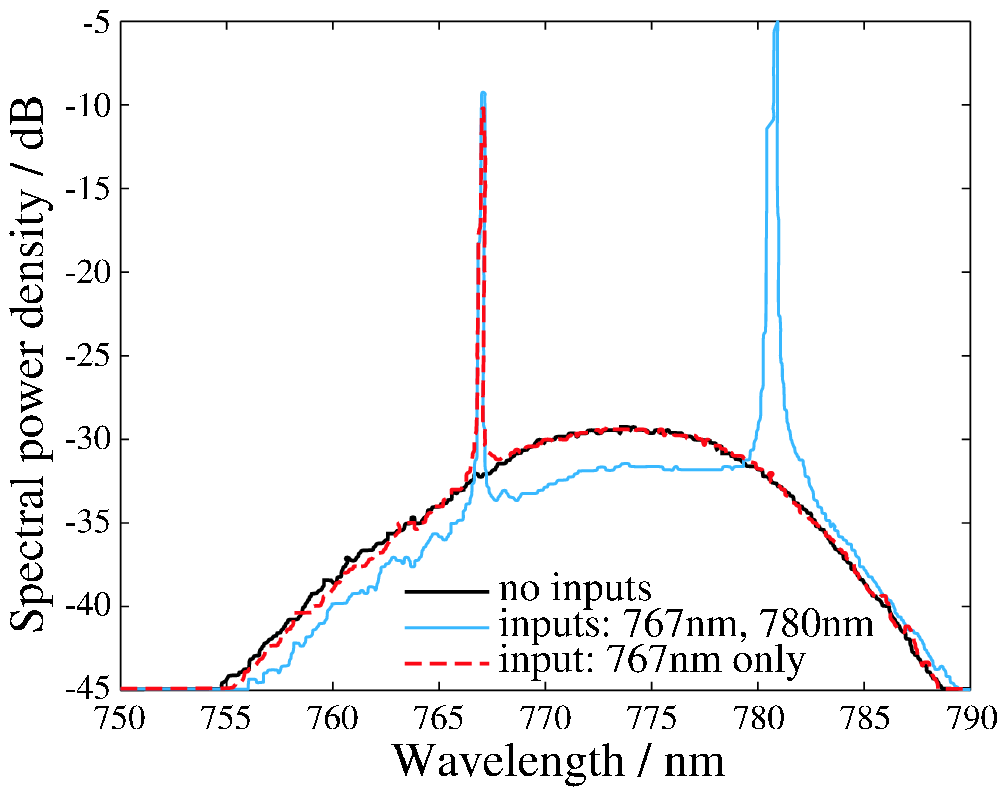}\\
  \caption{Output spectrum of tapered amplifier, after coupling to an
optical fibre. Two amplifiers were injected into the same fibre, one with
an MO at 767nm and one with an MO at 780nm, each amplifier was stabilised at a temperature
optimal for output power. Three curves show the total
ASE (black, solid), the amplified MO at 767nm with the ASE at 780nm (red,
dashed) and the sum of two amplified MOs at 767nm and 780nm (blue,
solid). Line widths and peak heights of amplified MOs are limited by the
detector resolution of 0.1nm. The ASE covers the region of high gain:
755--790nm.}\label{fig:mopa spectrum}
\end{figure}

The amplified spontaneous emission (ASE) of the OPA is indicative of the gain spectrum, typically extending between 755 and 790nm (\figref{fig:mopa spectrum}, thick black line). For 5mW of input power to an OPA, we observe around 400mW of output power (19dB amplification)\cite{LatestMOPA}. In this range the amplifier is below saturation intensity. However, when two beams of similar wavelength are injected into the same OPA, the output power is less than the sum of the outputs given when each beam is input individually (evidence of non-linear behaviour). When the same two beams used for measuring the linewidth of the MOs (\figref{fig:laser beats}) were injected, the amplified beat-note was not broadened: see \figref{fig:mopa beats}.

The output amplified light is beam-shaped (see \figref{fig:optical schema}) then injected into one or more fibres. Fibre-coupling efficiency can reach 60\%. The transmitted spectrum is shown in \figref{fig:mopa spectrum}. The typical transmitted spectral power density of the background amplified spontaneous emission (ASE) is less than 2mW per nm (equivalent to 4nW/MHz, or 24nW per $\Gamma$). For rubidium, with a beam of $1\,\text{cm}^2$ this corresponds to $10^4$ scattered photons per second per atom, for $200\,\text{mW}$ of useful light. Much of the suppression of unwanted light is due to spatial filtering, particularly on entry to the fibre: the ASE has a different divergence to the amplified injected light. This background light may have some unwanted effects on certain atomics physics experiments (eg coherent atomic manipulation) but will be of negligible effect when near resonance light is used (e.g. for laser-cooling experiments). The ASE can of course be cut off completely using shutters, when the amplified light is not desired.

We have found that the mechanical supports are very stable: injection MOs only need re-adjusting when the MO beam has been moved; fibre injection need to be re-optimised once per week at most. Once the input-focussing and output-collimation lenses are in place, they need never be moved.

\section{Conclusions}

We have presented a system of anti-reflection-coated diode lasers in extended cavities which can be tuned in the range 765--790nm at room temperature. Anti-reflection coating in conjunction with a carefully chosen grating helps reach shorter wavelengths and extract extra power. The light from these master oscillators can be amplified by nearly two orders of magnitude using tapered amplifiers, which are again suitable for both wavelengths. All parts are kept between 18 and 25$^\circ$C, so the mechanical designs are simple. The amplifiers have properties suitable for many atomic-physics experiments with rubidium and potassium: a narrow-band input gives a narrow-band output, the background noise spectrum is acceptably low intensity after spatial filtering by an optic fibre.

We have demonstrated the production of light suitable for experiments with atomic potassium, rubidium, or a mixture. Of particular interest is the fact that all of the electronics, mechanical mounts and thermal control elements are  home-made, thus significantly reducing the cost of building such complicated experiments.

We would like to thank Joseph Thywissen and David Gu\'{e}ry-Odelin for enlightening conversations, as well as Eagleyard and Opton Laser International for their semiconductors and after-sales service. RAN has been funded by the ESF BEC2000+ programme, and the EU Marie Curie training networks Cold Quantum Gases (FP5) and Atom Chips (FP6). This work is supported by CNES (DA:10030054), EU (grants IST-2001-38863, MRTN-CT-2003-505032 and FINAQS STREP), INTAS
(contract 211-855) and QUDEDIS.

\clearpage

\begin{figure}[p]
    \centering
    \includegraphics[angle=90,width=\textwidth]{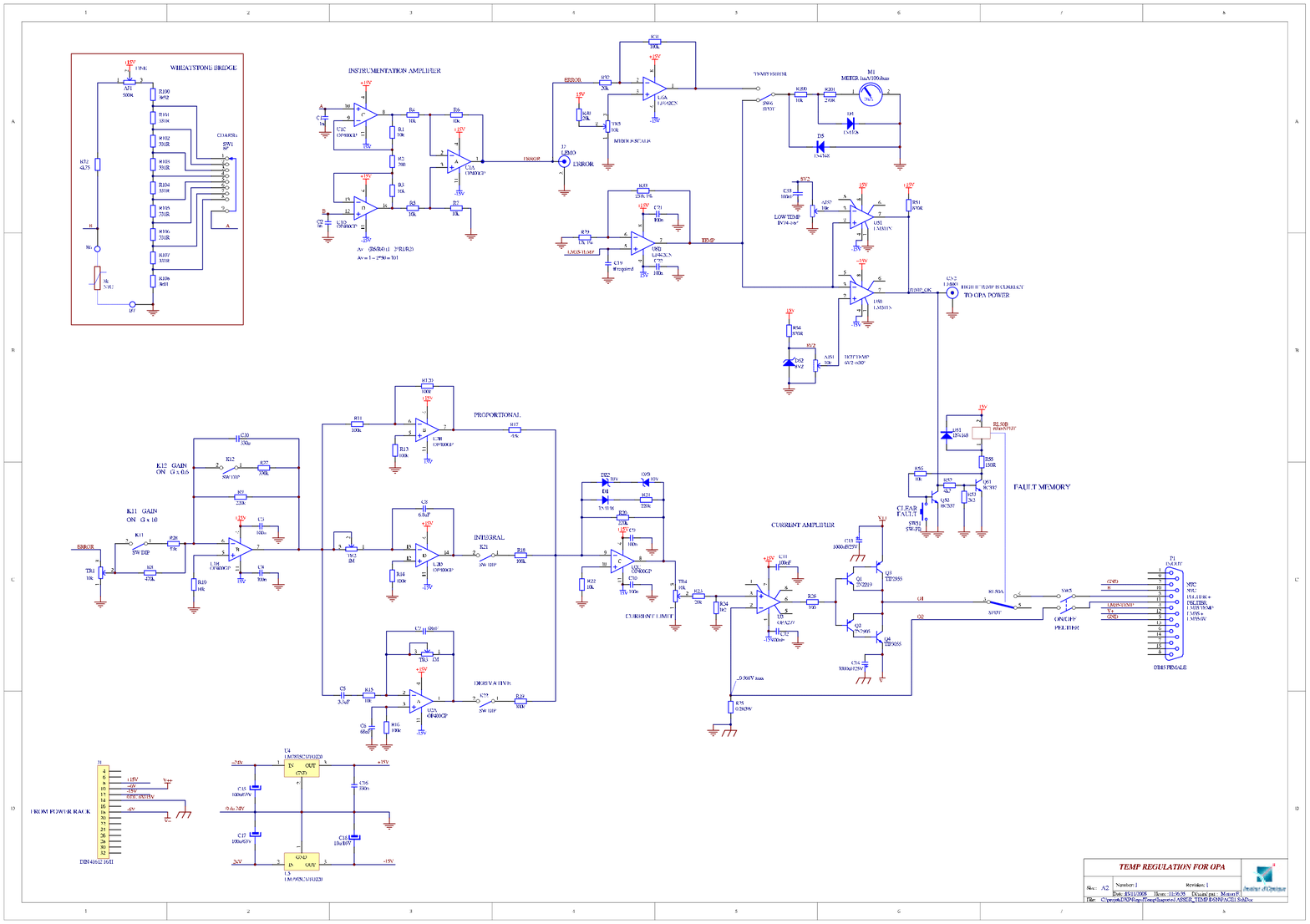}\\
    \caption{Electronic circuit diagram for the temperature control unit
for tapered amplifier.} \label{fig:elec mopa temperature}
\end{figure}

\clearpage

\begin{thebibliography}{1}
\bibitem{LatestMOPA} Tapered-amplifier chips are rapidly becoming more powerful. We have recently injected 3mW of light and acheived output of 600mW (23dB amplification) at 780nm. With the same chip we also managed fibre-coupling efficiency of 70\%, i.e. more than 400mW out of a monomode, polarisation-maintaining fibre.
\end{thebibliography}
\end{document}